\documentclass{PoS}
\usepackage{amsmath}
\usepackage{amssymb}
\usepackage{subfigure}

\newcommand{\g}[1]{\gamma_{#1}} % gamma matrices (covariant index)
\renewcommand{\l}{\left}
\renewcommand{\r}{\right}
\newcommand{\diag}{\mathrm{diag}}  % defines the diagonal-matrix abbr.
\newcommand{\chiral}[1]{\mathring{#1}} % places a "chiral" circle on top of a symbol

\newcommand{\order}[1]{\mathcal{O}\l({#1}\r)}

\newcommand{\SU}[1]{\mathrm{SU}\l(#1\r)}

\title{$\eta$ and $\eta'$ masses and decay constants from lattice QCD with $N_f=2+1+1$ quark flavours}

\ShortTitle{$\eta$ and $\eta'$ masses and decay constants from tmLQCD}

\author{\speaker{K.~Ottnad}\thanks{current affiliation:
    University of Cyprus, Nicosia, Cyprus}, C.~Urbach\\
  HISKP (Theory), University of Bonn, Nussallee 14-16, Bonn, Germany\\
  E-mail: \email{urbach@hiskp.uni-bonn.de}}

\author{Chris Michael\\
  Theoretical Physics Division, Department of Mathematical Sciences\\
  The University of Liverpool, Liverpool, L69 3BX, UK\\
  E-mail: \email{c.michael@liv.ac.uk}
}

\author{for the European Twisted Mass collaboration}

\abstract{We investigate the masses and decay constants of $\eta$ and
  $\eta'$ mesons using the Wilson 
  twisted mass formulation with $N_f=2+1+1$ dynamical quark flavours based on
  gauge configurations of ETMC. We show how to efficiently subtract
  excited state contributions to the relevant correlation functions and
  estimate in particular the $\eta'$ mass with improved precision. After
  investigating the strange quark mass dependence and the continuum and
  chiral extrapolations, we present our results for masses and mixing
  angle(s) at the physical point. Using chiral perturbation theory we
  also extract the decay constants $f_l$ and $f_s$ and use them to
  estimate the decay widths of $\eta,\eta'\to\gamma\gamma$ and the
  transition form factor in the limit of large momentum transfer.}

\FullConference{31st International Symposium on Lattice Field Theory - LATTICE 2013\\
		July 29 - August 3, 2013\\
		Mainz, Germany}

\begin{document}

\section{Introduction}

$\eta$ and $\eta'$ mesons are very interesting from a theoretical
point of view, because they are directly related to the $U_A(1)$
anomaly and topology in QCD. They are challenging to investigate
in lattice QCD  due to significant disconnected contributions. In a
series of papers and proceeding contributions we have presented results
for the corresponding meson masses and the mixing
angle(s)~\cite{Ottnad:2011mp,Ottnad:2012fv,Cichy:2012hq,Michael:2013gka}
using $N_f=2+1+1$ Wilson twisted mass fermions. In this proceeding we
extend our analysis  towards $\eta$ and $\eta'$ decay constants using
pseudoscalar matrix elements and chiral perturbation theory.

\begin{table}[t!]
 \centering
 \begin{tabular*}{1.\textwidth}{@{\extracolsep{\fill}}lcccccccc}
  \hline\hline
  ensemble & $\beta$ & $a\mu_l$ & $a\mu_\sigma$ & $a\mu_\delta$ & $L/a$ & $N_\mathrm{conf}$ & $N_s$ & $N_b$ \\ 
  \hline\hline
  $A30.32$   & $1.90$ & $0.0030$ & $0.150$  & $0.190$  & $32$ & $1367$ & $24$ & $5$  \\
  $A40.24$   & $1.90$ & $0.0040$ & $0.150$  & $0.190$  & $24$ & $2630$ & $32$ & $10$ \\
  $A40.32$   & $1.90$ & $0.0040$ & $0.150$  & $0.190$  & $32$ & $863$  & $24$ & $4$  \\
  $A60.24$   & $1.90$ & $0.0060$ & $0.150$  & $0.190$  & $24$ & $1251$ & $32$ & $5$  \\
  $A80.24$   & $1.90$ & $0.0080$ & $0.150$  & $0.190$  & $24$ & $2449$ & $32$ & $10$ \\
  $A100.24$  & $1.90$ & $0.0100$ & $0.150$  & $0.190$  & $24$ & $2493$ & $32$ & $10$ \\
  \hline
  $A80.24s$  & $1.90$ & $0.0080$ & $0.150$  & $0.197$  & $24$ & $2517$ & $32$ & $10$ \\
  $A100.24s$ & $1.90$ & $0.0100$ & $0.150$  & $0.197$  & $24$ & $2312$ & $32$ & $10$ \\
  \hline
  $B25.32$   & $1.95$ & $0.0025$ & $0.135$  & $0.170$  & $32$ & $1484$ & $24$ & $5$  \\
  $B35.32$   & $1.95$ & $0.0035$ & $0.135$  & $0.170$  & $32$ & $1251$ & $24$ & $5$  \\
  $B55.32$   & $1.95$ & $0.0055$ & $0.135$  & $0.170$  & $32$ & $1545$ & $24$ & $5$  \\ 
  $B75.32$   & $1.95$ & $0.0075$ & $0.135$  & $0.170$  & $32$ & $922$  & $24$ & $4$  \\ 
  $B85.24$   & $1.95$ & $0.0085$ & $0.135$  & $0.170$  & $24$ & $573$  & $32$ & $2$  \\
  \hline
  $D15.48$   & $2.10$ & $0.0015$ & $0.120$  & $0.1385$ & $48$ & $1045$ & $24$ & $10$ \\
  $D30.48$   & $2.10$ & $0.0030$ & $0.120$  & $0.1385$ & $48$ & $469$  & $24$ & $3$  \\
  $D45.32sc$ & $2.10$ & $0.0045$ & $0.0937$ & $0.1077$ & $32$ & $1887$ & $24$ & $10$ \\
  \hline\hline
  \vspace*{0.1cm}
 \end{tabular*}
 \caption{The ensembles used in this investigation. For the labelling
   we employ the notation of ref.~\cite{Baron:2010bv}. Additionally,
   we give the number of configurations $N_\mathrm{conf}$, the number
   of stochastic samples $N_s$ for all ensembles and the bootstrap
   block length $N_b$. The D30.48 ensemble was not yet included in
   Ref.~\cite{Ottnad:2012fv}.}
 \label{tab:setup}
\end{table}

The results we present are based on gauge configurations provided by
the European Twisted Mass Collaboration (ETMC) and correspond to three
values of the lattice spacing, $a=0.061$ fm, $a=0.078$ fm and
$a=0.086$ fm. The pion masses range from $230$ to $500$
MeV~\cite{Baron:2010bv,Baron:2010th,Baron:2011sf}. A list of the
investigated ensembles is given in Table~\ref{tab:setup}. For setting
the scale we use throughout this proceeding contribution the Sommer
parameter $r_0 = 0.45(2)\ \mathrm{fm}$~\cite{Baron:2011sf}.

We use the Wilson twisted mass formulation of lattice
QCD~\cite{Frezzotti:2000nk,Frezzotti:2003xj} with the main advantage
of automatic $\mathcal{O}(a)$ improvement at maximal
twist~\cite{Frezzotti:2003ni} and the disadvantage that 
parity and flavour symmetry are both broken at finite values of the
lattice spacing. The latter was shown to
affect mainly the value of the neutral pion
mass~\cite{Urbach:2007rt,Dimopoulos:2009qv,Baron:2009wt}. Furthermore,
for the non-degenerate quark doublet this introduces mixing between
charm and strange quarks. For details on the lattice action we refer
to Ref.~\cite{Baron:2010bv}.

\section{Pseudoscalar flavour-singlet mesons}

We compute the Euclidean correlation functions
\begin{equation}
  \label{eq:correlations}
  \mathcal{C}(t)_{qq'} =
  \langle\mathcal{O}_q(t'+t)\mathcal{O}_{q'}(t')\rangle\,,\quad
  q,q'\in{l,s,c}\,,  \end{equation} with operators $\mathcal{O}_l =
(\bar ui\gamma_5 u + \bar d i\gamma_5 d)/\sqrt{2}$, $\mathcal{O}_s =
\bar s i\gamma_5 s$ and $\mathcal{O}_c = \bar c i\gamma_5 c$. We enlarge
our correlator matrix $\mathcal{C}$ by including also fuzzed operators.
Note that in twisted mass lattice QCD there are several steps required
to reach these correlation functions, as explained in detail in
Ref.~\cite{Ottnad:2012fv}. We estimate the disconnected contributions to
the correlation functions Eq.~\ref{eq:correlations} using Gaussian
volume sources and  the connected contributions~\cite{Boucaud:2008xu}
using the one-end trick. For the light disconnected contributions a
powerful noise reduction technique is
available~\cite{Jansen:2008wv,Ottnad:2012fv}. For the strange and charm
disconnected loops, we use the hopping parameter noise reduction
technique~\cite{Boucaud:2008xu}.

We solve the generalised eigenvalue problem
(GEVP)~\cite{Michael:1982gb,Luscher:1990ck,Blossier:2009kd}
\begin{equation}
  \label{eq:GEVP}
  \mathcal{C}(t)\eta^{(n)}(t,t_0) =
  \lambda^{(n)}(t,t_0)\mathcal{C}(t_0)\eta^{(n)}(t,t_0) 
\end{equation}
% to diagonalise $\mathcal{C}$ with 
for eigenvalues $\lambda^{(n)}(t,t_0)$
and eigenvectors $\eta^{(n)}$. $n$ labels the states $\eta,
\eta^\prime,...$ contributing. Masses of these states can be determined
from the exponential fall-off of $\lambda^{(n)}(t,t_0)$ at large $t$.
The pseudoscalar matrix elements $A_{q,n}\equiv\langle
n|\mathcal{O}_q|0\rangle$ with $q \in l,s,c$ and $n\in
\eta,\eta',...$ can be extracted from the
eigenvectors~\cite{Blossier:2009kd}. It turns out that the charm quark
contributions to $\eta,\eta'$ are negligible and, thus, we drop
the $c$ quark in what follows.

\subsection*{Decay Constants and Mixing}

In general, decay constants are defined for any pseudoscalar meson
$\mathrm{P}$ from axial vector matrix elements 
\begin{equation}
 \l<0\r| A^a_\mu \l|\mathrm{P}\l(p\r) \r> = i f^a_\mathrm{P} p_\mu \,,
\label{eq:decay_constant_definition}
\end{equation}
which leads to
\begin{equation}
 \l<0\r| \partial^\mu A^a_\mu \l|\mathrm{P}\l(0\r) \r> = f^a_\mathrm{P} M_\mathrm{P}^2 \,,
\end{equation}
for projection to zero momentum. Assuming that $\eta$ and $\eta'$ are
not flavour eigenstates, each of them exhibits a coupling to the
singlet and octet axial vector current $A^0_\mu$ and $A^8_\mu$,
respectively. Therefore, one ends up with four independent decay
constants for the $\eta$,$\eta'$-system, which are commonly
parametrised in terms of two decay constants $f_0$, $f_8$ and two
mixing angles $\theta_0$, $\theta_8$ 
\begin{equation}
  \l( \begin{array}{ll}
    f_\eta^8 & f_\eta^0 \\
    f_{\eta'}^8 & f_{\eta'}^0 
  \end{array}\r) = \l( \begin{array}{rr}
    f_8 \cos \theta_8  & -f_0 \sin \theta_0 \\
    f_8 \sin \theta_8 & f_0 \cos \theta_0
  \end{array} \r) \equiv \Xi\l(\theta_8, \theta_0\r) \diag \l(f_8,\, f_0\r) \,.
  \label{eq:octet_singlet_basis_parametrisation}
\end{equation}
The singlet decay constant $f_0$ needs renormalisation, determined by
the anomalous dimension of the axial singlet
current~\cite{Kodaira:1979pa}. The dependence 
on the scale is $\mathcal{O}(1/N_C)$ and can, therefore, be dropped in
the following discussion. For a detailed discussion see
Refs.~\cite{Kaiser:1998ds,Kaiser:2000gs}. 

On the lattice it is more convenient to work in the quark flavour
basis, with the axial vector currents $A_\mu^0$ and $A_\mu^8$ replaced
by the combinations 
\begin{alignat}{2}
 A^l_\mu =& \frac{2}{\sqrt{3}} A^0_\mu + \sqrt{\frac{2}{3}} A^8_\mu =&& \frac{1}{\sqrt{2}} \l(\bar{u} \g{\mu} \g{5} u + \bar{d} \g{\mu} \g{5} d\r) \,, \label{eq:quark_flavor_basis_1} \\ 
 A^s_\mu =& \sqrt{\frac{2}{3}} A^0_\mu - \frac{2}{\sqrt{3}} A^8_\mu =&& \bar{s} \g{\mu} \g{5} s \,, \label{eq:quark_flavor_basis_2}
\end{alignat}
in which the light quarks and the strange quark contributions are
disentangled. In exact analogy to the singlet-octet basis this basis
again allows for a parametrisation in terms of two decay constants and
two mixing angles 
\begin{equation}
 \l( \begin{array}{ll}
      f_\eta^l & f_\eta^s \\
      f_{\eta'}^l & f_{\eta'}^s 
     \end{array}\r) = \Xi\l(\phi_l,\phi_s\r)  \diag\l(f_l,\, f_s\r) \,,
 \label{eq:quark_flavor_basis_parametrisation}
\end{equation}
where the mixing matrix $\Xi$ has the same form as the one defined in
Eq.~\ref{eq:octet_singlet_basis_parametrisation}. In this basis it
is also expected that we
have~\cite{Schechter:1992iz,Kaiser:1998ds,Kaiser:2000gs,Feldmann:1998sh,Feldmann:1998vh}
\begin{equation}
 \l| \frac{\phi_l - \phi_s}{\phi_l + \phi_s}\r| \ll 1 \,.
\end{equation}
motivating a simplified mixing scheme in the quark flavour basis with
only one angle $\phi$ 
\begin{equation}
  \l( \begin{array}{ll}
    f_\eta^l & f_\eta^s \\
    f_{\eta'}^l & f_{\eta'}^s 
  \end{array}\r) = \Xi\l(\phi\r) \diag\l(f_l,\, f_s\r) + \order{\Lambda_1} \,,
  \label{eq:simplified_mixing_scheme}
\end{equation}
where $\Xi\l(\phi\r) \equiv \Xi\l(\phi, \phi\r)$ and $\Lambda_1$
parametrises residual OZI violating terms. The mixing angle $\phi$ is
related to the double ratio of amplitudes 
\begin{equation}
 \tan^2\l(\phi\r) = -\frac{f_{\eta'}^l f_\eta^s}{f_\eta^l f_{\eta'}^s}
 \,. \label{eq:phi_double_ratio_axial_vector} 
\end{equation}
Axial vector current matrix elements turn out to be difficult to
measure in actual simulations due to noise. This is why it is most
convenient to consider pseudoscalar currents in the quark flavour basis
in analogy to
Eqs.~(\ref{eq:quark_flavor_basis_1}),(\ref{eq:quark_flavor_basis_2}). 
\begin{align}
 P^l &= \frac{1}{\sqrt{2}} \l( \bar{u} \g{5} u + \bar{d} \g{5} d \r) \,, \label{eq:pseudoscalar_quark_flavor_basis_1} \\
 P^s &= \bar{s} i\g{5} s \,, \label{eq:pseudoscalar_quark_flavor_basis_2}
\end{align}
such that the matrix elements for pseudoscalar mesons $\mathrm{P}$ are given by
\begin{equation}
 h_\mathrm{P}^i = 2 m_i \l<0\r| P^i \l|\mathrm{P}\r> \,,
 \label{eq:pseudoscalar_matrix_elements}
\end{equation}
which are free from renormalisation.
Making use of $\chi$PT and dropping subleading terms leads to the
following expression~\cite{Feldmann:1999uf}
\begin{equation}
  \l( \begin{array}{ll}
    h_\eta^l & h_\eta^s \\
    h_{\eta'}^l & h_{\eta'}^s 
  \end{array}\r) = \Xi\l(\phi\r) \diag\l(M_\pi^2 f_l,\, \l(2 M_\mathrm{K}^2 - M_\pi^2\r) f_s\r)\,.
  \label{eq:pseudoscalar_quark_flavor_basis_parametrisation}
\end{equation}
This expression allows access to the decay constants $f_s$ and
$f_l$ from pseudoscalar matrix elements under the assumption that
$\chi$PT can be applied. In terms of pseudoscalar matrix elements the
mixing angle $\phi$ is obtained as
\begin{equation}
 \tan^2\l(\phi\r) = -\frac{h_{\eta'}^l h_\eta^s}{h_\eta^l h_{\eta'}^s}
 \,, \label{eq:phi_double_ratio_pseudo_scalar} 
\end{equation}
where actually quark masses and renormalisation constants drop out in
the ratio. Expanding again to two angles, $\phi_l$ and $\phi_s$ are
written as
\begin{equation}
  \label{eq:phils_ratio_pseudo_scalar}
  \tan(\phi_l) = \frac{h_{\eta'}^l}{h_\eta^l}\,,\qquad \tan(\phi_s) = -
  \frac{h_{\eta}^s}{h_{\eta'}^s}\,. 
\end{equation}
Finally we remark that in order to compute the matrix elements $h_P^i$
in Wilson twisted mass lattice QCD the ratio of renormalisation
constants $Z_P/Z_S$ is required~\cite{Ottnad:2012fv} which we took
from Ref.~\cite{ETM:2011aa}.

\section{Excited State Removal}

\begin{figure}[t]
  \centering
  \subfigure[]{\includegraphics[width=.48\linewidth]
    {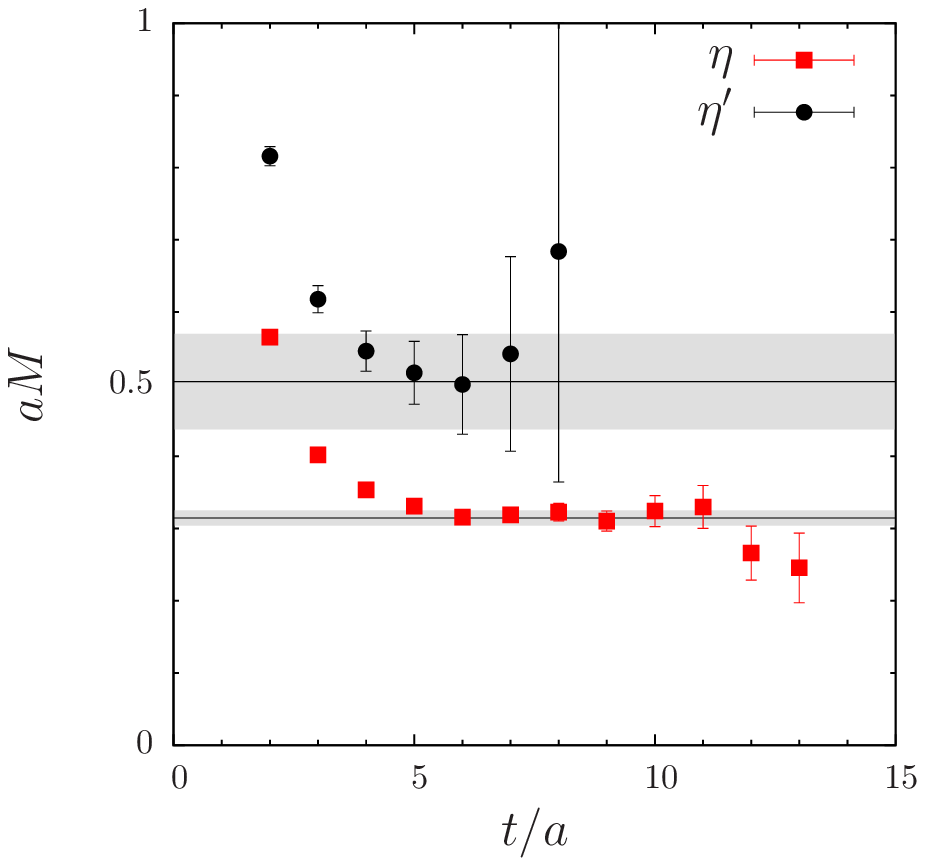}}\quad
  \subfigure[]{\includegraphics[width=.48\linewidth]
    {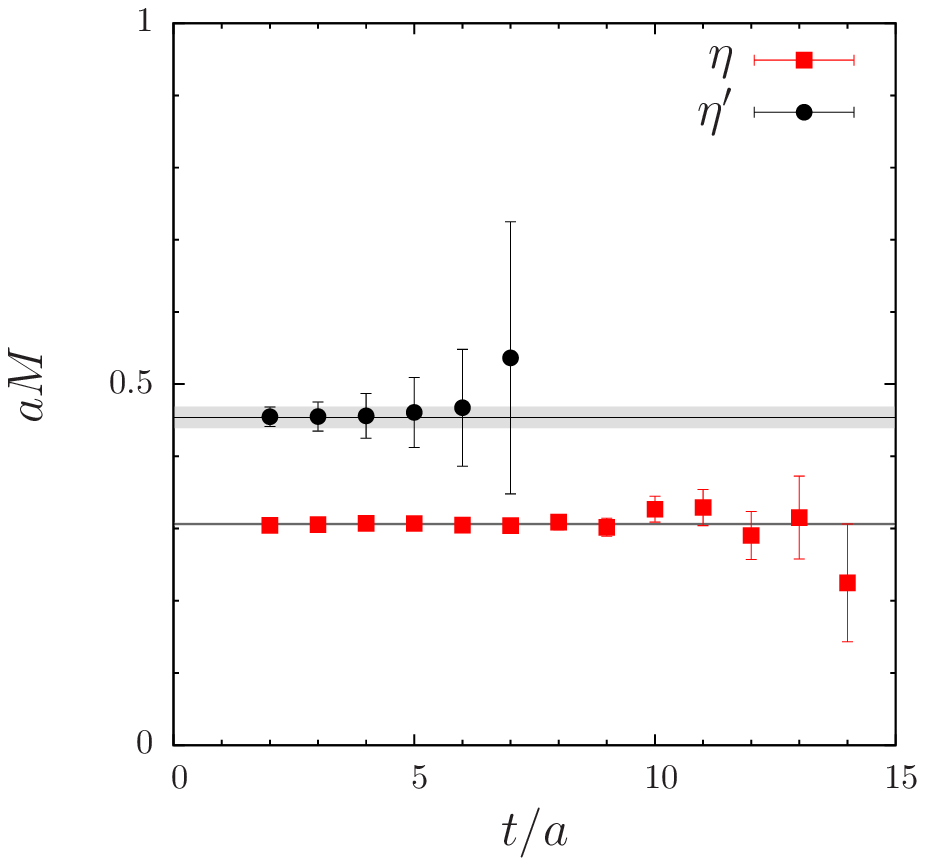}}\quad
  \caption{(a) Effective masses in lattice units determined from solving
    the GEVP for a $6\times6$ matrix with $t_0/a=1$ for ensemble
    A100. (b) the same, but after removal of excited states in the
    connected contributions.}
  \label{fig:effmass}
\end{figure}

The result of solving the GEVP for a $6\times6$ matrix for ensemble
A100 including fuzzed operators is shown as effective masses in the
left panel of figure~\ref{fig:effmass}. It is visible that the ground
state, the $\eta$ meson, can be extracted with good precision, while
for the $\eta'$ meson it is unclear that a plateau is reached before
the signal is lost in noise.

However, there is a possibility to obtain a significant
improvement for the extraction of $\eta'$ mass (and further
observables) using a powerful method to separate ground and excited
states, which has first been proposed in \cite{Neff:2001zr} and that
has already been successfully employed for the case of the $\eta_2$
for two dynamical quark flavours in \cite{Jansen:2008wv}. In the
following we will describe this method and apply it to our data.

The method is based on the assumption that the quark disconnected
diagrams give a sizeable contribution only to the $\eta$ and $\eta'$
state, but are negligible for any heavier state with the same quantum
numbers. Considering the fluctuations of the topological charge which
are expected to give a dominant contribution to the mass of the
$\eta'$, this assumption would be valid if these fluctuation mainly
couple to the $\eta$ and $\eta'$ states. Still, the validity of this
assumption needs to be carefully checked from our data and may
introduce systematic uncertainties.

Since the quark connected contributions exhibit a constant
signal-to-noise ratio, it is in principle possible to determine the
respective ground states at sufficiently large $t/a$ with high
statistical accuracy and without any significant contamination from 
higher states. After fitting the respective ground states of the
connected correlators, we can use it to subtract the excited state
contributions such that the full connected correlators are replaced by
correlators that contain only the ground state. Note that for sufficiently large
$t/a$ this reproduces the original ground state by construction.

Now, if the aforementioned assumption holds, i.e. the disconnected
diagrams are relevant only to the two lowest states $\eta$, $\eta'$
one should obtain a plateau in the effective mass at very low values
of $t/a$ after solving the GEVP.  The result of the procedure is shown
in the right panel of
figure~\ref{fig:effmass}. Indeed, one observes a
plateau for both states starting basically at the lowest possible
value of $t/a$. Furthermore, a comparison with the effective masses
from the standard $6\times6$ matrix in the left panel of
figure~\ref{fig:effmass} reveals that the plateau
values agree very well within their respective errors. Most
importantly, the data in the right panel allows for a much better
accuracy in the determination of both masses as the point errors are
much smaller at such low values of $t/a$. Therefore, we will use this
method for all the results presented in this proceeding contribution. 

However, we remark that in the twisted mass formulation with the
non-degenerate doublet this procedure is in practice restricted to the
connected correlation functions corresponding to physical light and
strange quarks. This is due to the violation of flavour symmetry in the
heavy sector of the twisted mass formulation, implying that the four
connected contributions in the heavy sector will all yield the same  
ground state. This ground state corresponds to an artificial particle,
i.e. a connected-only, neutral pion-like particle made out of strange
quarks. Therefore, we will restrict ourselves in the following
discussion to the analysis of a $2\times2$ correlation function matrix
corresponding to (local) physical operators made of light and strange
quarks.

\section{$\eta$ and $\eta'$ Masses and Extrapolations}

\begin{figure}[!t]
 \centering
 \subfigure[]{\includegraphics[width=.445\linewidth]{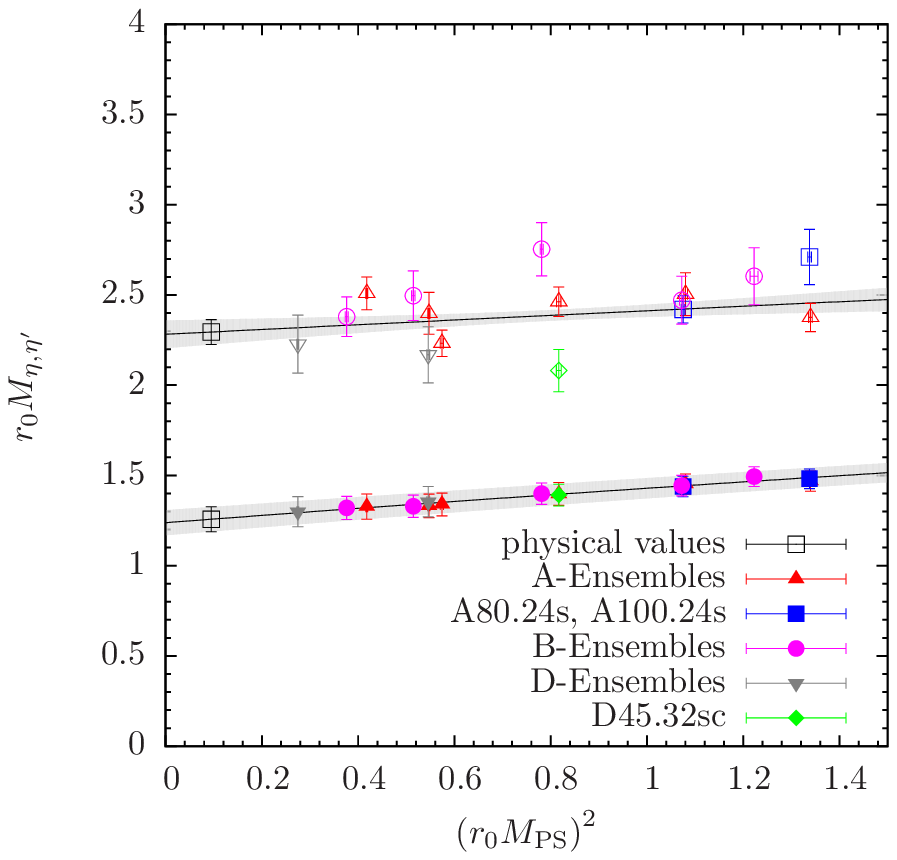}}\quad
 \subfigure[]{\includegraphics[width=.445\linewidth]{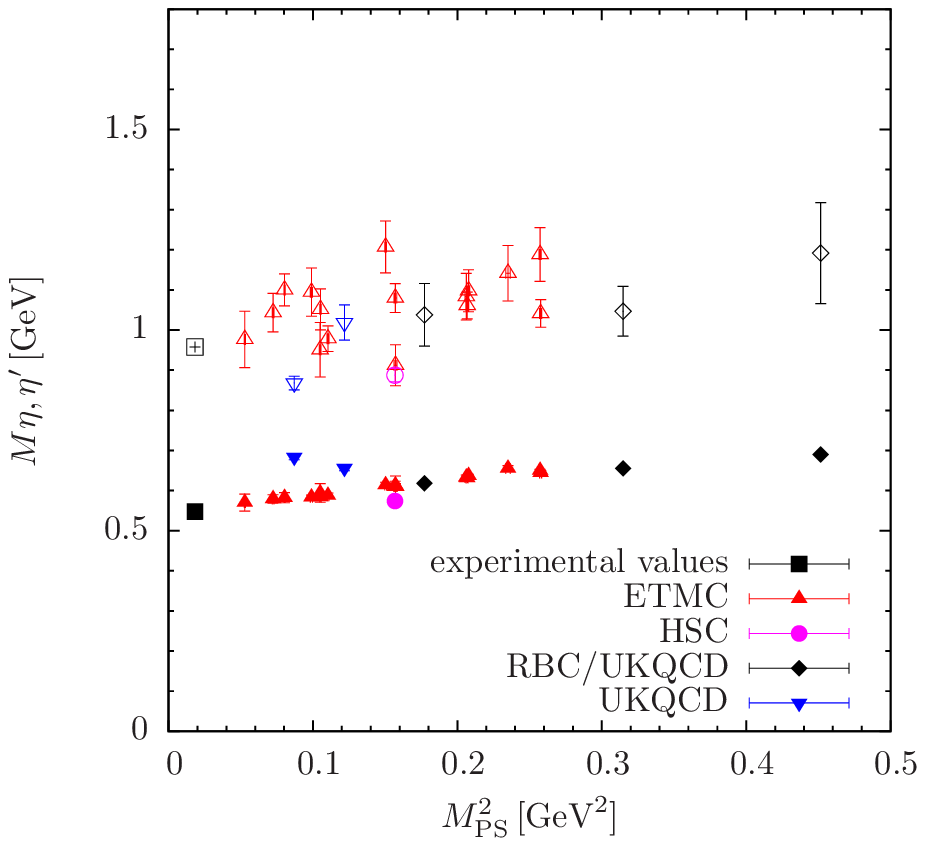}}
 \caption{(a) Our results for $r_0 \overline{M}_\eta$ (filled symbols)
   (corrected for the mismatch in $r_0 M_\mathrm{K}$) and $r_0 M_{\eta'}$ (open symbols). The fitted 
   curves are linear functions in $(r_0 M_\mathrm{PS})^2$ as discussed
   in the text. (b) The same data as in (a), but in physical units and
   including results from other lattice computations.} 
 \label{fig:eta_trick_M_i_tuned}
\end{figure}

We determine the masses using the excited state removal method
described in the previous section for all ensembles listed in
table~\ref{tab:setup} and determine $aM_\eta$ and $aM_{\eta'}$. It
turns out that in $aM_\eta$ we see a strange quark mass dependence
that we can resolve well within our statistical errors. As the physical
values of the strange quark masses vary by about 10\% in between the
different lattice spacing values, we have to correct for this. This is
described in detail in Ref.~\cite{Ottnad:2012fv}. Here we will
repeat the procedure only shortly: we use the ensembles A80 and A80s
(A100 and A100s), which differ in the bare strange quark mass value,
to estimate $D_\eta = dM_\eta^2/dM_K^2$. Next we correct all $aM_\eta$
values to correspond to a line of $M_K[M_\mathrm{PS}^2]$ values which
reproduces the physical kaon mass at
$M_\mathrm{PS}=M_{\pi^0}^\mathrm{phys}$. These corrected values we denote 
 with $\overline{M}_\eta$. Note that for this procedure we ignore any
dependence of $D_\eta$ on the quark masses and the lattice spacing.
For the $\eta'$ mass we do not resolve quark mass or lattice spacing
dependence within our errors, so we do not attempt to correct for
those. 

The results for $\overline{M}_\eta$ and $M_{\eta'}$ are
summarised in figure~\ref{fig:eta_trick_M_i_tuned}, where we show
$r_0\overline{M}_\eta$ as filled and $M_{\eta'}$ as open symbols,
respectively, both as functions of $(r_0M_\mathrm{PS})^2$. For both
mesons the data fall within errors on a single line such that in both
cases we model the data for $(r_0M)^2$ as constant plus a linear term
in $(r_0M_\mathrm{PS})^2$. The corresponding best fit to the data and 
 its error band is shown as lines with shaded bands. The error band for
$M_\eta$ is mainly due to the error of $D_\eta$.

After extrapolating to the physical point and converting to physical
units we obtain
\begin{equation}
 M_\eta(M_\pi) = 551(11)_\mathrm{stat}(6)_\mathrm{sys} \,\mathrm{MeV} \,,
 \label{eq:eta_trick_M_eta_phys}
\end{equation}
where the systematic error has been estimated from fitting to the data at
each value of the lattice spacing separately. Note that the value for
the physical $\eta$ mass is in very good agreement with 
the experimental value $M_\eta^\mathrm{exp}=547.85(2)\
\mathrm{MeV}$~\cite{Beringer:1900zz}.
In addition, for the $\SU{2}$ chiral
limit we find $r_0\chiral{M}_\eta^{\SU{2}} =
1.24(7)_\mathrm{stat}(2)_\mathrm{sys}$, which yields 
\begin{equation}
 \chiral{M}_\eta^{\SU{2}} = 543(11)_\mathrm{stat}(7)_\mathrm{sys} \,\mathrm{MeV} \,.
\end{equation}
We may extrapolate further quantities in order to check the
validity of our correction procedure for mistuned values of the
strange quark mass. First, we consider the GMO ratio
determined directly from the data
and perform an extrapolation in $\l(r_0
M_\mathrm{PS}\r)^2$. However, it turns out that taking the uncorrected
values of $M_\eta$ the extrapolation misses the experimental value
$(3M_\eta^2/(4M_\mathrm{K}^2 - M_\pi^2))^\mathrm{exp} = 0.925$
considering only the statistical error by more than $2\sigma$, i.e. we
obtain $\left(3M_\eta^2/(4M_\mathrm{K}^2 - M_\pi^2)\right)_{M_\pi} =
0.963(15)_\mathrm{stat}(35)_\mathrm{sys}$. This may be seen as a hint
that the significantly increased statistical precision of the improved
analysis strategy allows to resolve a residual strange quark mass
dependence which is not cancelled in the ratio. This was not
possible with the statistical accuracy we could obtain in
Ref.~\cite{Ottnad:2012fv}. Note that compared to
the direct extrapolation the statistical precision is even further
enhanced for the case of dimensionless ratios because the physical
value of $r_0$ is only required for fixing the physical point, but not
for the conversion to physical units. Considering only the
$B$ ensembles for which the value of the strange quark mass is close
to physical yields $\left(3M_\eta^2/(4M_\mathrm{K}^2 -
  M_\pi^2)\right)_{M_\pi}^B = 0.928(27)_\mathrm{stat}$, indicating
that the systematic error is mainly caused by such a residual
effect. Indeed, using the corrected values $\bar{M}_\eta$ and the
corresponding values of the kaon mass to calculate the GMO ratio the
extrapolation gives $\left(3M_\eta^2/(4M_\mathrm{K}^2 -
  M_\pi^2)\right)_{M_\pi} = 0.946(26)_\mathrm{stat}(22)_\mathrm{sys}$,
which agrees nicely with the experimental value and exhibits a smaller
systematic error compared to the result obtained from using the
uncorrected values of $M_\eta$. For the physical value of the $\eta$
mass we obtain
\begin{equation}
 M_\eta = 554(8)_\mathrm{stat}(7)_\mathrm{sys} \ \mathrm{MeV}\,,
 \label{eq:eta_trick_M_eta_phys_GMO}
\end{equation}
in agreement with the result from the direct extrapolation and the
experimental value.

A similar picture arises from the light quark mass extrapolation of
the ratio $M_\eta/M_K$.
Using the uncorrected values of $M_\eta$ which
yields $\l(M_\eta/M_\mathrm{K}\r)_{M_\pi}=
1.117(8)_\mathrm{stat}(23)_\mathrm{sys}$, missing the experimental
value $\l(M_\eta/M_\mathrm{K}\r)^\mathrm{exp} = 1.100$ again by
roughly $2\sigma$ if taking only the statistical error into
account.
Like for the case of the GMO ratio
taking only the $B$ ensembles into account gives nice agreement with
the experimental value, i.e. $\l(M_\eta/M_\mathrm{K}\r)_{M_\pi}^B=
1.117(8)_\mathrm{stat}$, whereas $A$ and $D$ ensembles give
significantly larger values. Therefore, we have also repeated the
extrapolation using the corrected values $\bar{M}_\eta$ and the
corresponding kaon masses.
Within the statistical errors one
again obtains excellent agreement with experiment,
i.e. $\l(M_\eta/M_\mathrm{K}\r)_{M_\pi}=
1.099(16)_\mathrm{stat}(19)_\mathrm{sys}$, which results in 
\begin{equation}
 M_\eta = 0.547(8)_\mathrm{stat}(9)_\mathrm{sys} \ \mathrm{MeV} \,,
\end{equation}
compatible with the results from direct and GMO ratio extrapolations.

In order to obtain our final result for the physical mass of the
$\eta$ we take the weighted average from the three previously
discussed methods used for the extrapolation in the light quark
mass. Accounting for any correlations, this yields 
\begin{equation}
 M_\eta = 551(8)_\mathrm{stat}(6)_\mathrm{sys}\,.
 \label{eq:final_M_eta_phys}
\end{equation}
which is in excellent agreement with experiment and exhibits
substantially smaller errors compared to Ref.~\cite{Ottnad:2012fv}.

For the $\eta'$ we obtain
\begin{equation}
 M_{\eta'} = 1006(54)_\mathrm{stat}(38)_\mathrm{sys}(+61)_\mathrm{ex} \,,
\label{eq:}
\end{equation}
from a linear extrapolation in $(r_0M_\mathrm{PS})^2$ of all the
(uncorrected) data.
In order to quantify a possible error introduced by the excited
state removal in the connected contributions, we quote the difference
between the extrapolations with and without excited state removal as
an additional systematic error. Again, the standard systematic error
has been determined from fits to the data at single values of the
lattice spacing and it is interesting to note that it actually turns
out to be the smallest of the three errors. Within the larger errors
this result is again in very good agreement with experiment,
confirming that QCD indeed accounts for the significantly larger mass
of the $\eta'$ that is observed experimentally. 

In the right panel of figure~\ref{fig:eta_trick_M_i_tuned} we show a
compilation of our results for $\eta$ and $\eta'$ masses
together with results available in the literature for
$N_f=2+1$ flavour lattice QCD. For $M_\eta$ we show the values
corrected for the mismatch in $M_\mathrm{K}$. We remark that in
\cite{Kaneko:2009za} $\eta$ and $\eta'$ meson masses have been
computed using $N_f=2+1$ flavours of overlap quarks at one value of the
lattice spacing and large values of the pion mass, however, in this
reference not enough details are given to be included in our
comparison figure~\ref{fig:eta_trick_M_i_tuned}. The results
in \cite{Christ:2010dd} have been obtained using $N_f=2+1$ flavours of
domain wall fermions and again for a single value of the lattice
spacing $a\approx 0.1\,\mathrm{fm}$ but for three values of the pion
mass in a range from $\sim 400\, \mathrm{MeV}$ to $\sim 700\,
\mathrm{MeV}$. The corresponding data points in
figure~\ref{fig:eta_trick_M_i_tuned} are labelled
``RBC/UK        QCD''. Another single data point is added from
\cite{Dudek:2011tt} by the Hadron Spectrum Collaboration (HSC) for
which Wilson fermions have been employed. Again, it was not possible
to include more recent results by the HSC \cite{Dudek:2013yja} due to
the lack of explicit numerical values in this reference for the
relevant masses. Finally, in \cite{Gregory:2011sg} data from staggered
fermions are presented for two different values of the lattice spacing
with each of them also at a different value of the pion mass. In
figure~\ref{fig:eta_trick_M_i_tuned} the corresponding data
points are labelled ``UKQCD''. The figure suggests an overall agreement
between all collaborations.

\section{$\eta$, $\eta'$-mixing Angles} \label{sec:mixing}

\begin{figure}[!t]
 \centering
 \subfigure[]{\includegraphics[height=.445\linewidth]{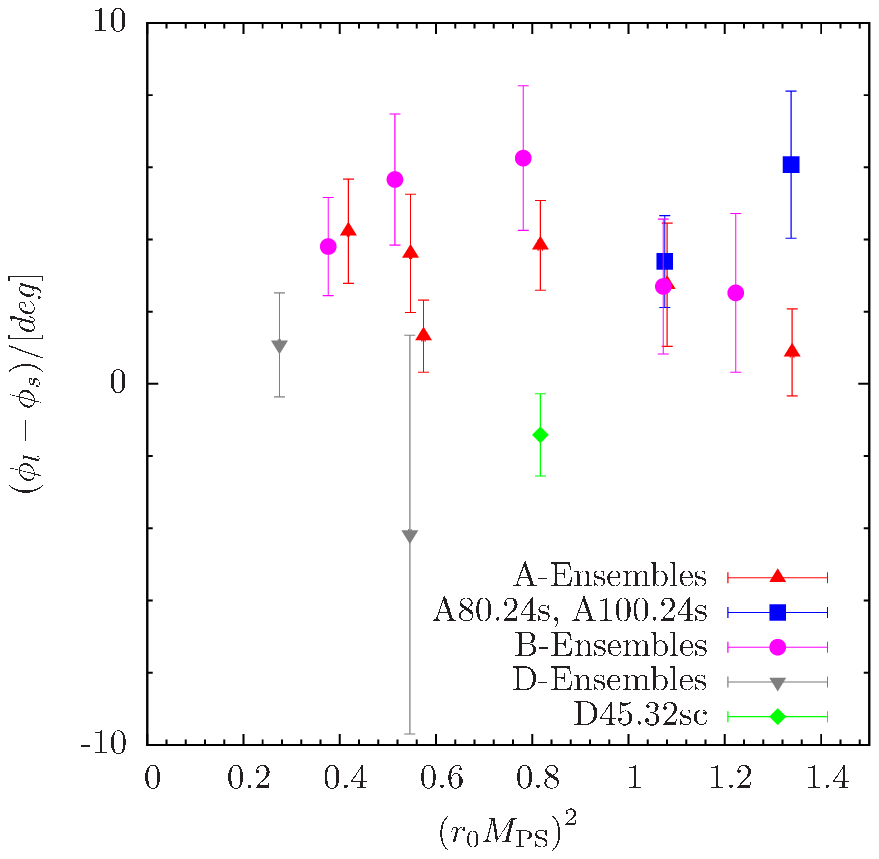}}\quad
 \subfigure[]{\includegraphics[height=.445\linewidth]{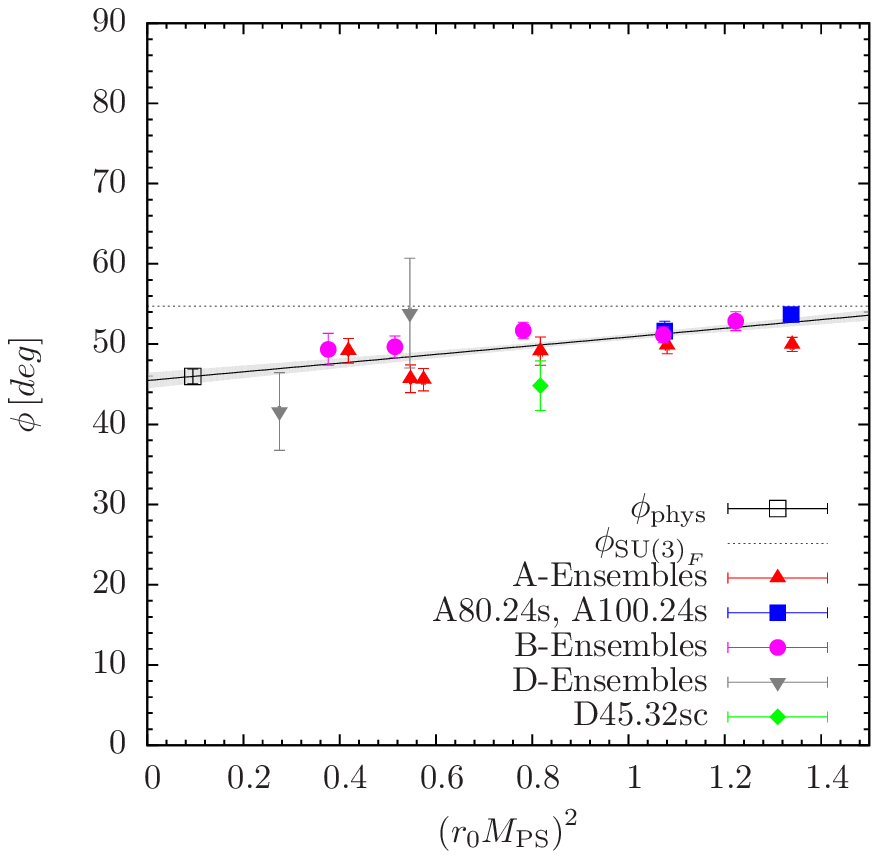}}
 \caption{(a) the difference $\phi_l-\phi_s$ in degrees and (b) $\phi$
 as functions of $(r_0M_\mathrm{pS})^2$.}
 \label{fig:angles}
\end{figure}

The mixing angles $\phi_l$ and $\phi_s$ can be extracted using
Eq.~\ref{eq:phils_ratio_pseudo_scalar} with a mixing model based on
the pseudoscalar matrix elements $h^q_P$. In the left panel of
figure~\ref{fig:angles} we show $\phi_l-\phi_s$ in degrees as a
function of $(r_0M_\mathrm{PS})^2$. One observes that this quantity is
consistent with zero within errors. Also an extrapolation to the
physical point yields $3(1)_\mathrm{stat}(3)_\mathrm{syst}^\circ$,
where the systematic error is estimated from the maximal difference
compared to extrapolating the data sets for the three different
lattice spacings separately.

In the right panel of figure~\ref{fig:angles} we show the average
angle $\phi$ (Eq.~\ref{eq:phi_double_ratio_pseudo_scalar}) in degrees
as a function of $(r_0M_\mathrm{PS})^2$, with 
smaller statistical errors than $\phi_l$ and $\phi_s$ separately,
because of correlation in the matrix elements. Our precision is
not sufficient to resolve any residual lattice spacing or strange
quark mass dependence. Hence, we extrapolate linearly in
$(r_0M_\mathrm{PS})^2$ and obtain
\begin{equation}
 \phi = 46(1)_\mathrm{stat}(3)_\mathrm{sys}^\circ \,,
 \label{eq:eta_trick_phi_phys}
\end{equation}
where the first error is statistical and the second systematic from
fitting the three values of the lattice spacing separately. 

Besides the mixing angle $\phi$, we consider the angles $\phi_l$,
$\phi_s$ which are relevant to cross-check the assumptions entering
our mixing scheme. Again, we have
performed linear fits in $\l(r_o M_\mathrm{PS} \r)^2$
and obtain at the physical value of the pion mass
\begin{equation}
 \phi_l = 48(1)_\mathrm{stat}(4)_\mathrm{sys}^\circ \,, \qquad
 \phi_s = 44(1)_\mathrm{stat}(3)_\mathrm{sys}^\circ \,, 
 \label{eq:eta_trick_phi_l_phi_s_phys}
\end{equation}
where the systematic uncertainties have been determined in the same
way as for the angle $\phi$ itself. The results are compatible  within
errors. Notably, for $\phi_s$ there is very good
agreement for the results within each of the two set ($A80.24$,
$A80.24s$) and ($A100.24$, $A100.24s$), indicating that the influence
of the strange quark is smaller for this quantity and in general more
of the data points lie within the error band of the linear
fit.

\section{Decay Constants}

\begin{figure}[!t]
 \centering
 \subfigure[]{\includegraphics[height=.445\linewidth]{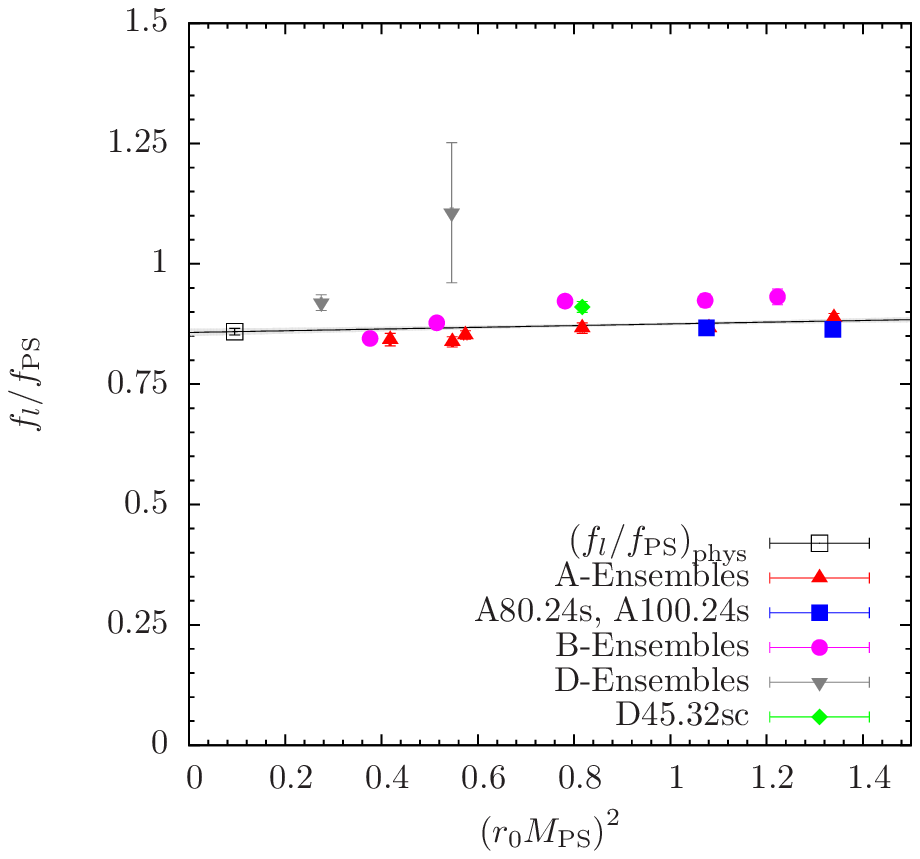}}\quad
 \subfigure[]{\includegraphics[height=.445\linewidth]{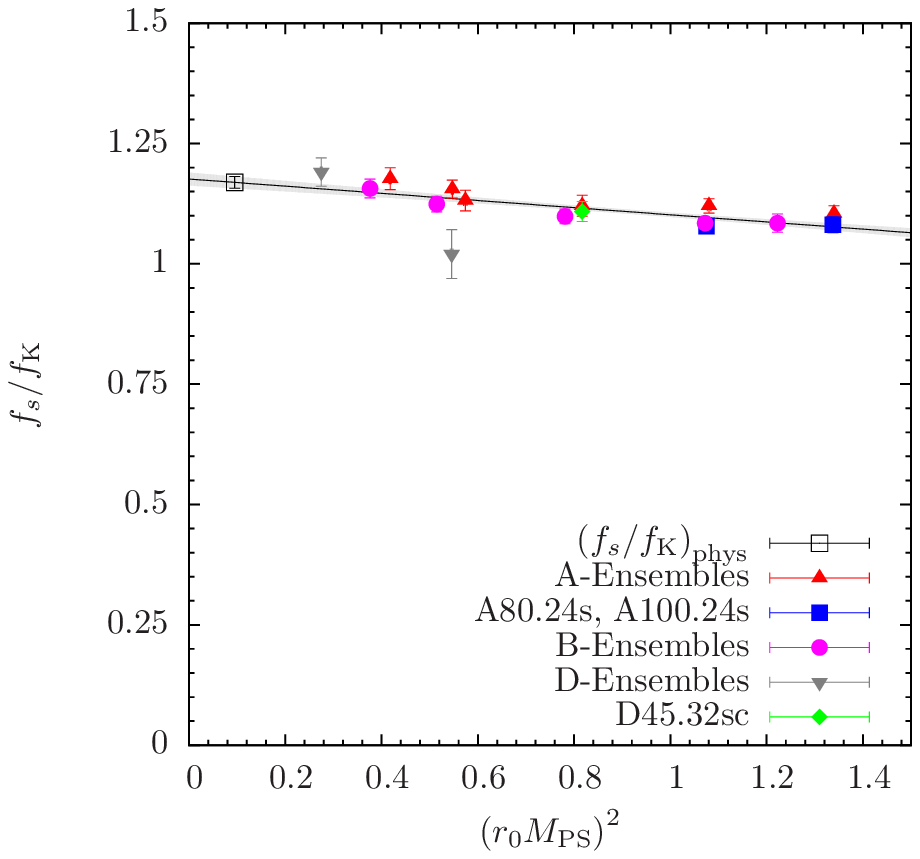}}
 \caption{(a) $f_l/f_\mathrm{PS}$ and (b) $f_l/f_K$ as functions of $(r_0M_\mathrm{PS})^2$.}
 \label{fig:decay_constants}
\end{figure}

The application of the excited state removal method discussed in the
previous section allows to extract the decay constants $f_l$
and $f_s$ to a rather high statistical precision by means of the
pseudoscalar matrix elements $h_P^i$ and using
Eq.~\ref{eq:pseudoscalar_quark_flavor_basis_parametrisation}. Of
course, one needs to keep in mind that this is based on the assumption
that the underlying chiral perturbation theory analysis is
valid. Currently, we cannot estimate a corresponding systematic
uncertainty.

In figure~\ref{fig:decay_constants} we show $f_l/f_\mathrm{PS}$ and
$f_s/f_K$ as functions of $\l(r_0 M_\mathrm{PS} \r)^2$ in the left and
right panel, respectively. We have chosen to plot these ratios because
it appears that most of the quark mass and lattice spacing
dependence cancels. A linear extrapolation of $f_l/f_\mathrm{PS}$ in
$\l(r_0 M_\mathrm{PS} \r)^2$ to the physical point results in
\begin{equation}
 f_l/f_\pi = 0.859(7)_\mathrm{stat}(64)_\mathrm{sys} \,.
 \label{eq:eta_trick_f_l_over_f_PS}
\end{equation}
However, from the plot it appears that there is still a rather sizeable
dependence on the lattice spacing present while the strange quark mass
dependence seems to cancel in ratio as all $A$-ensembles fall on one
single curve. In fact, the systematic error estimated from fitting the
data at each value of the lattice spacing separately is one order of
magnitude larger than the statistical error and there is clear trend
towards larger values of $f_l/f_\mathrm{PS}$ extrapolated to the
physical pion mass for decreasing values of the lattice
spacing. Therefore, we additionally quote the result of a linear fit
restricted to the data at the finest lattice spacing, which yields
\begin{equation}
  \label{eq:fl}
 \l(f_l/f_\pi \r)^D= 0.924(22)_\mathrm{stat}\,. 
\end{equation}

For the ratio $f_s/f_K$ most of the strange quark mass dependence is
cancelled and the data seem almost perfectly linear in the light quark
mass, exhibiting only a moderate slope. Moreover, for this case there
are no discernible scaling artefacts within errors and the data are
well described by a linear fit which gives 
\begin{equation}
  f_s/f_K = 1.166 (11)_\mathrm{stat}(31)_\mathrm{sys} \,,\qquad
  f_s/f_\pi = 1.336(13)_\mathrm{stat}(37)_\mathrm{sys}\,,
 \label{eq:eta_trick_f_s_over_f_K}
\end{equation}
at the physical value of the pion mass. Clearly, the systematic error
is significantly smaller than the one obtained for the physical value
of $f_l/f_\pi$, confirming the smallness of any residual
lattice artefacts or strange quark mass dependence for $f_s/f_K$. 
The values we obtain are in rough agreement to phenomenological
values~\cite{Feldmann:1999uf} (and for a very recent one see
Ref.~\cite{Escribano:2013kba}), because the spread in the
phenomenological estimates is quite large.
Still, our estimate for $f_l/f_\pi$ is a bit lower than
expected. We are investigating ways to better control systematics in
our analysis.

Finally we have to remark that finite volume effects might play an
important role for the decay constants, but they hopefully cancel in
the ratios we used.

As discussed in Ref.~\cite{Escribano:2013kba}, this determination of
mixing parameters can be used to better estimate the hadronic
light-by-light contribution to the anomalous magnetic moment of the
muon. Of course, eventually a computation of the corresponding
transition form factors is desired as recently performed for the
neutral pion in Ref.~\cite{Feng:2012ck}.

% this translates to a value for $f_l/f_s = ?$ which can be used 
% to compute the ratio of gluon matrix elements!x

\section{Decay Widths $\Gamma_{P\to\gamma\gamma}$}

\begin{figure}[!t]
 \centering
 \subfigure[]{\includegraphics[height=.4\linewidth]{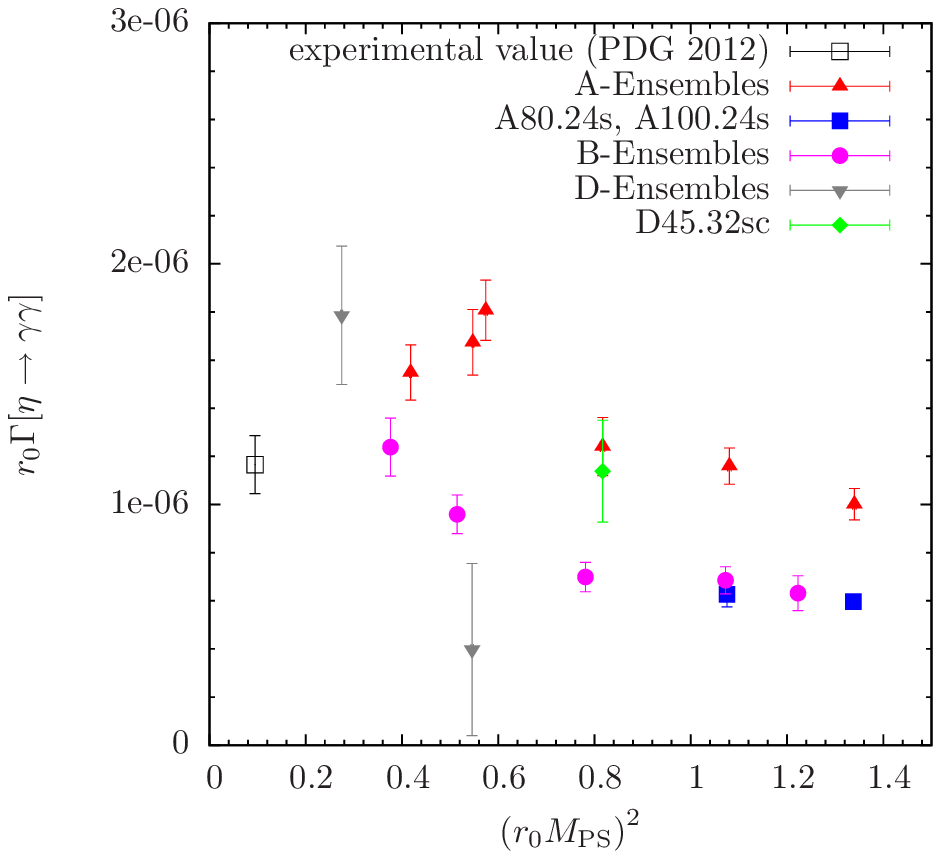}}\quad
 \subfigure[]{\includegraphics[height=.4\linewidth]{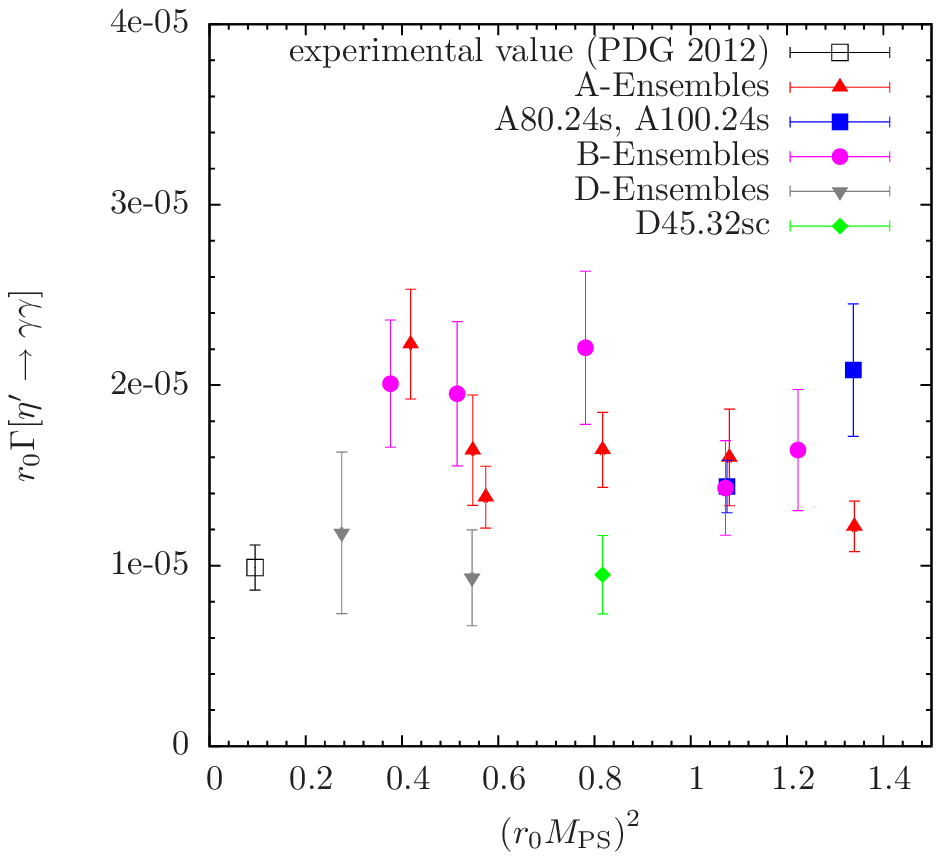}}
 \caption{We show the decays widths $\Gamma_{\eta\to\gamma\gamma}$ (a)
   and $\Gamma_{\eta'\to\gamma\gamma}$ (b) in units of $r_0$ as
   functions of $(r_0 M_\mathrm{PS})^2$. In addition we show the
   corresponding estimate from the PDG~\protect{\cite{Beringer:1900zz}}.}
 \label{fig:decay_width}
\end{figure}

The decay constants $f_l$ and $f_s$ are important low energy
constants. However, they can also be used to estimate
phenomenologically interesting quantities, most prominently the decay
widths of $\eta,\eta'\ \to\ \gamma\gamma$. To the same order in the
effective theory one can relate the decays widths with the mixing
parameters in the quark flavour basis as
follows~\cite{Feldmann:1997vc,Feldmann:2002kz} (see
Refs.~\cite{Kaiser:1998ds,Kaiser:2000gs} for how to include these
quantities into the effective field theory framework)
\begin{equation}
  \label{eq:widths}
  \begin{split}
    \Gamma_{\eta\to\gamma\gamma}\ &=\
    \frac{\alpha^2}{288\pi^3}M_\eta^3\left[\frac{5}{f_l}\cos\phi -
      \frac{\sqrt{2}}{f_s}\sin\phi\right]^2\,,\\
    \Gamma_{\eta'\to\gamma\gamma}\ &=\
    \frac{\alpha^2}{288\pi^3}M_{\eta'}^3\left[\frac{5}{f_l}\sin\phi +
      \frac{\sqrt{2}}{f_s}\cos\phi\right]^2\,,\\
  \end{split}
\end{equation}
where again OZI suppressed terms have been dropped and our
normalisation is that $f_\pi=130.7\ \mathrm{MeV}$. Using our lattice
data for $M_{\eta,\eta'}$, $\phi$, $f_l$ and $f_s$, we have computed
the decay widths and show them in units of the Sommer parameter as a
function of $(r_0 M_\mathrm{PS})^2$ in
figure~\ref{fig:decay_width}. We also include the PDG value for
convenience~\cite{Beringer:1900zz}. For
$\Gamma_{\eta'\to\gamma\gamma}$ we observe a flat dependence of the
light quark mass and agreement  between our data and the PDG value
within our errors. However, the data show a tendency for lower values
with decreasing values of the lattice spacing. 

The situation is not so clear for $\Gamma_{\eta\to\gamma\gamma}$, where
the lattice data is of the right magnitude compared to the PDG
value. But there are clearly strange quark mass and light quark mass
effects visible that we cannot control at the moment. In addition
there might be lattice artefacts.

Two remarks are in order: first of all the mass dependence of the
widths has not been computed in effective field theory. Hence, we have
no rigorous means to extrapolate our data to the physical
point. Eqs.~\ref{eq:widths} are strictly speaking only valid in the
chiral limit. Second, the PDG value for $\Gamma_{\eta\to\gamma\gamma}$
does not include Primakoff experiments, which give a significantly
smaller value\footnote{We thank P.~Masjuan for pointing our attention
  to this fact.}.

Related quantities are the pseudoscalar transition form
factors $F_{\eta\gamma\gamma^*}$ and $F_{\eta'\gamma\gamma^*}$ for
large momentum transfer $q^2$, which can be 
expressed as~\cite{Escribano:2013kba}
\begin{equation}
  \label{eq:tff_asym}
  \begin{split}
  \lim_{q^2\to\infty}q^2 F_{\eta\gamma\gamma^*}(q^2) &= \frac{10}{3\sqrt{2}}
  f_l \cos\phi - \frac{2}{3}f_s\sin\phi\ =\
  96(5)_\mathrm{stat}(25)_\mathrm{sys}\ \mathrm{MeV}\,,\\
  \lim_{q^2\to\infty}q^2 F_{\eta'\gamma\gamma^*}(q^2) &= \frac{10}{3\sqrt{2}}
  f_l \sin\phi + \frac{2}{3}f_s\cos\phi\ =\ 274(3)_\mathrm{stat}(11)_\mathrm{sys}\ \mathrm{MeV}\,.\\
  \end{split}
\end{equation}
These results have to be understood to be very preliminary. They are
obtained by using the values of $f_l$ and $f_s$ Eqs.~\ref{eq:fl} and
\ref{eq:eta_trick_f_s_over_f_K}, respectively, and the one for $\phi$ from
Eq.~\ref{eq:eta_trick_phi_phys}. The systematic uncertainty is again
calculated from the maximal difference to results from the separate
lattice spacing values.

\section{Summary}

We have presented results for $\eta$ and $\eta'$ masses and mixing
parameters from lattice QCD with $2+1+1$ dynamical quark flavours. The
computation is based on gauge configurations provided by the ETM
collaboration. Due to an efficient excited state removal method 
we could determine $M_\eta$, $M_{\eta'}$ and the
mixing angles to good accuracy. For the masses we find excellent
agreement with experiment. The mixing angles in the quark flavour basis
confirm that $|\phi_l - \phi_s|/|\phi_l+\phi_s|\ll 1$ and we find the
single angle to be close to $46$ degrees. This indicates that the
$\eta'$ is dominantly a flavour singlet state.

For the first time we present results for the decay constants $f_l$
and $f_s$ using chiral perturbation theory. We find similar values to those
found in phenomenology. It is important to keep in mind that $f_l$ and
 $f_s$ are likely to be affected by significant systematic
uncertainties due to residual lattice artefacts, quark mass dependence
and the approximation in chiral perturbation theory we use.

The extraction of $f_l$ and $f_s$ gives us the unique opportunity
to estimate the decay widths of $\eta\to\gamma\gamma$ and
$\eta'\to\gamma\gamma$. Despite the fact that we do not have a
rigorous formula for the extrapolation of our data to the physical
point we observe ballpark agreement of our data with the current PDG 
estimate. 

We thank all members of ETMC for the most enjoyable
collaboration. The computer time for this project was made
available to us by the John von Neumann-Institute for
Computing (NIC) on the JUDGE and Jugene systems. In
particular, we thank U.-G. Mei{\ss}ner 
for granting us access on JUDGE. We thank P.~Masjuan for helpful
discussions. This project was
funded by the DFG as a project in the SFB/TR 16. K.~O.
and C.~U. were supported by the BCGS of Physics and
Astronomie. The open source software packages
tmLQCD~\cite{Jansen:2009xp,urbachcode:lat2013,bartekcode:lat2013},
Lemon~\cite{Deuzeman:2011wz}, and R~\cite{R:2005} have been used.

\bibliographystyle{h-physrev5}
\bibliography{bibliography}

\end{document}